\documentclass[
reprint,
superscriptaddress,
amsmath,amssymb,
aps,
pre,
]{revtex4-2}
\usepackage[colorlinks=true,linkcolor=blue,citecolor=blue,urlcolor=blue]{hyperref}
\usepackage{orcidlink}
\providecommand{\p}{\partial}
\providecommand{\dod}[3][]{\mathinner{\dfrac{\mathrm{d}^{#1}#2}{\mathrm{d}^{\vphantom{#1}}{{#3}^{#1}}}}}

\providecommand{\dpd}[3][]{\mathinner{\dfrac{\p^{#1}#2}{\p^{\vphantom{#1}}{{#3}^{#1}}}}}
\providecommand{\dfd}[3][]{\dfrac{\delta{^{#1}}#2}{\delta^{{#1}}{{#3}^{#1}}}}
\newcommand{\abs}[1]{\lvert {#1} \rvert}
\newcommand{\grad}[2][]{\vec{\nabla}\!{#1}{#2}}
\newcommand{\divj}[2][]{\vec{\nabla}\!{#1}\cdot{#2}}
\newcommand{\lapl}[1][]{\vec{\nabla}^2_{\mkern-4mu{#1}\mkern3mu}}
\providecommand{\intj}[4][]{\int_{#2}^{#1} \! {#3} \, \mathrm{d} {#4}}

\begin{document}

\title{Nonequilibrium thermodynamic foundation of the grand-potential phase field model}
\author{Jin Zhang\,\orcidlink{0000-0003-4982-1830}}\email{jzhang@northwestern.edu}
\affiliation{Department of Materials Science and Engineering, Northwestern University, Evanston, Illinois 60208, USA}
\author{James A. Warren}
\affiliation{Material Measurement Laboratory, National Institute of Standards and Technology, 100 Bureau Drive, Gaithersburg, Maryland 20899, USA}
\author{Peter W. Voorhees}
\affiliation{Department of Materials Science and Engineering, Northwestern University, Evanston, Illinois 60208, USA}

\begin{abstract}
Choosing the correct free energy functional is critical when developing thermodynamically consistent phase field models. We show that the grand-potential phase field model minimizes the Helmholtz free energy when mass conservation is imposed. While both functionals are at a minimum at equilibrium, the Helmholtz free energy decreases monotonically with time in the grand-potential phase field model, whereas the grand potential does not. Minimizing the grand potential implies a different problem where a system can exchange mass with its surroundings at every point, leading to a condition of isochemical potential and invalidating mass conservation of the system.
\end{abstract}

\maketitle

The phase field method is widely used to describe the evolution of nonequilibrium systems. One of its most broad and successful applications is modeling the solidification of alloys. Early phase field models for alloys \cite{Wheeler1992} do not decouple the bulk and interfacial energies, so they rely on a very thin interface (typically on the nanometer scale) to provide quantitative predictions. This imposes a significant computational constraint for practical applications of the models to systems ranging from micrometers to millimeters in size. The Kim-Kim-Suzuki (KKS) model \cite{Kim1999} addressed this problem by viewing the interface as a two-phase mixture, allowing for a natural decoupling of the bulk and interfacial contributions, and permitting accurate predictions with significantly larger interface widths. However, to achieve this, a local equilibrium constraint must be fulfilled. This constraint, which involves the equality of the diffusion potentials for the two phases ($\tilde{\mu}^\alpha = \tilde{\mu}^\beta$), is a nonlinear equation for general free energies and must be solved at each grid point and each timestep, a computationally expensive undertaking. To avoid this, carefully crafted free energy densities can be used. Plapp \cite{Plapp2011} proposed a grand-potential formulation by switching from the concentration $c$ to the diffusion potential $\tilde{\mu}$ as the natural variable. This treatment naturally fulfills the local equilibrium constraint and fully decouples the bulk and interfacial energies. This model has been extended to multiphase and multicomponent systems \cite{Choudhury2012,Aagesen2018} and is widely used for modeling alloy solidification and various other applications \cite{Cogswell2015,Zhang2023}.

Though not explicitly mentioned in the original paper \cite{Plapp2011}, many researchers have assumed that the grand-potential phase field model minimizes the system's total grand potential since the grand potential was used in Plapp's approach \cite{KubendranAmos2020,Khanna2020,Chatterjee2021,Xuan2022,Sessim2022,Boutin2022,Mamaev2023,Xue2024,Muntaha2024,Bayle2020,Verdier2024}. Some researchers have pointed out that the evolution equations from the grand-potential phase field model do not appear to result from a variational derivation of the system's total grand potential, but the point has not been emphasized \cite{Danilov2014,Chatterjee2023,Zhang2023}.

We show here that the grand-potential phase field model proposed by Plapp \cite{Plapp2011} actually minimizes the Helmholtz free energy under proper boundary conditions. Here, ``minimize'' pertains to nonequilibrium thermodynamics: the free energy not only reaches an extremum at equilibrium, as described in classical thermodynamics, but also decreases monotonically with time (energy dissipation or the energy being a Lyapunov functional \cite{Langer1986,Penrose1990}). Moreover, the concept of ``equilibrium'' should be viewed as a state where all fast processes have happened while all slow processes have not, as highlighted by Feynman \cite{Feynman2018}. For example, solidification of an alloy can often be approximated as a process that takes place in \textit{thermal equilibrium}, a state with rapid heat diffusion and slow mass diffusion. Thus, in the isothermal case, heat exchange is fast and temperature remains uniform while mass diffusion exists as a nonequilibrium process. The Helmholtz free energy decreases monotonically in this process. The case where the total grand potential is monotonically decreasing is related to the presence of \textit{chemical equilibrium} at a timescale significantly longer than mass diffusion or a system with infinite diffusivity. An example analogous to isothermal is isochemical potential (or iso-$\tilde{\mu}$), where mass exchange with the surroundings is fast to maintain the iso-$\tilde{\mu}$ condition, leading to the lack of mass conservation in the \textit{system}. Note that not all phase field models are designed to conserve mass. As Plapp uses mass conservation in his model, the system is not under an iso-$\tilde{\mu}$ condition, and the total grand potential does not decrease monotonically in time, despite the grand potential reaching its minimum at equilibrium (when mass diffusion is done). This observation in no way renders Plapp's model incorrect, but this misapprehension could lead to substantive errors in interpreting the free energy functional and constructing thermodynamically consistent phase field models.

Let us briefly review Plapp's grand-potential phase field model \cite{Plapp2011}. The grand potential of a binary alloy system is defined as
\begin{equation}\label{eq:GP:Omega:binary}
  \Omega[\phi,\tilde{\mu}] = \int_{V}\! \frac{1}{2}\kappa \abs{\grad{\phi}}^2 + m g(\phi) + \omega(\phi,\tilde{\mu})  {\,\mathrm{d}V},
\end{equation}
where $\phi$ is the phase field variable, $\tilde{\mu}$ is the diffusion potential, $\kappa$ is the gradient coefficient, $g(\phi)$ is the double-well function, $m$ is the height of the well, and the bulk grand-potential density is interpolated from the two phases
\begin{equation}
  \label{eq:GP:omegad:binary}
  \omega=p(\phi) \omega^\alpha(\tilde{\mu}) + (1-p(\phi)) \omega^\beta(\tilde{\mu}),
\end{equation}
where $p(\phi)$ is an interpolation function, and $\omega^\alpha$ and $\omega^\beta$ are the grand-potential densities of each phase, which are both functions of the diffusion potential $\tilde{\mu}$.\\
The functional derivatives of the system's grand potential with respect to the two independent variables are
\begin{align}\label{eq:GP:dOmega}
  \dfd{\Omega}{\phi} &= - \kappa \lapl{\phi} + m g'(\phi) + p'(\phi) (\omega^\alpha - \omega^\beta),\\
  -\dfd{\Omega}{\tilde{\mu}} &= p(\phi) c^\alpha + (1-p(\phi)) c^\beta = c,
\end{align}
where $c$ is the total concentration interpolated from the phase concentrations $c^\alpha = - {\partial\omega^\alpha}/{\partial \tilde{\mu}}$.\\
The evolution of the phase field variable $\phi$ follows the Allen-Cahn equation
\begin{equation}
  \label{eq:GP:evolution:phi:binary}
  \dpd{\phi}{t} = - M_\phi \dfd{\Omega}{\phi},
\end{equation}
where $M_\phi$ is the phase field mobility and is positive. \\
The evolution of the diffusion potential $\tilde{\mu}$ is derived from the mass conservation equation
\begin{equation}
  \label{eq:GP:evolution:c:binary}
  \dpd{c}{t}  = -\divj{\vec{J}} - j,
\end{equation}
where $\vec{J}$ is the mass flux and $j$ is the body mass flux. We now consider the case without a source term $j=0$. In the grand-potential phase field model, Eq.~\ref{eq:GP:evolution:c:binary} is reformulated to an evolution equation of the diffusion potential
\begin{equation}
  \label{eq:GP:evolution:mu:binary}
  \chi \dpd{\tilde{\mu}}{t} = \divj{\left(M \grad{\tilde{\mu}}\right)} - p'(\phi) (c^\alpha-c^\beta) \dpd{\phi}{t},
\end{equation}
where $\chi=\partial c/\partial \tilde{\mu}$ is the susceptibility, $p'(\phi)=\mathrm{d}p/\mathrm{d}\phi$, and we use
\begin{equation}
  \label{eq:GP:evolution:J}
\vec{J}= -M \grad{\tilde{\mu}},
\end{equation}
where $M$ is an atomic mobility which is positive.

To determine if the total grand potential is monotonically decreasing in time, we take the time derivative of the system's grand potential [Eq.~\ref{eq:GP:Omega:binary}]
\begin{align}\label{eq:GP:dOmegadt}
  \dod{\Omega}{t}&= \intj{V}{\dfd{\Omega}{\phi}\dpd{\phi}{t} + \dfd{\Omega}{\tilde{\mu}}\dpd{\tilde{\mu}}{t}}{V}\nonumber\\
                 &= \intj{V}{\dfd{\Omega}{\phi}\dpd{\phi}{t} - c \dpd{\tilde{\mu}}{t}}{V} \leq 0.
\end{align}
Neglecting cross-coupling between $\phi$ and $\tilde{\mu}$, for the grand potential to decrease in time requires the integrand to be negative-definite, which, when combined with Eq.~\ref{eq:GP:evolution:phi:binary}, yields the condition
\begin{align}
-M_\phi \left(\dfd{\Omega}{\phi}\right)^2  -c \dpd{\tilde{\mu}}{t} \leq 0. \label{eq:GP:dOmegadt:mu}
\end{align}
For this to hold for any virtual variation in $\phi$ or $\tilde{\mu}$, each term must itself be negative \cite{Truesdell1965}. From Eq.~\ref{eq:GP:evolution:mu:binary}, the second term can be written as $-c \partial_t\tilde{\mu}=-c \partial_t c/\chi$ for regions far from the boundary ($\partial_t \phi=0$). Since the concentration is nonnegative $c\geq 0$, this necessitates $c$ to monotonically increase/decrease with time for a positive/negative $\chi$. However, this is not always the case, at least in a subdomain of the system. It is evident that the evolution equations of the grand-potential phase field model [Eqs.~\ref{eq:GP:evolution:phi:binary} and \ref{eq:GP:evolution:mu:binary}] do not minimize the total grand potential [Eq.~\ref{eq:GP:Omega:binary}]. In other words, $\mathrm{d}\Omega/\mathrm{d}t$ is not guaranteed to be negative. For some problems, like order-disorder transformations, mass conserved is not required by the dynamics. In other cases, like nucleation, the small size of the nucleus ensures a constant $\tilde{\mu}$. Equation~\ref{eq:GP:dOmegadt:mu} is applicable in both scenarios but not for any processes involving diffusion.

Thus, the model does not appear to guarantee that the grand potential energy decreases during evolution. However, as will be shown below, it dissipates the Helmholtz free energy. The Helmholtz free energy of the system is
\begin{equation}\label{eq:F:binary}
  F[\phi,c] = \int_{V}\! \frac{1}{2}\kappa \abs{\grad{\phi}}^2 + m g(\phi) + f(\phi,c) {\,\mathrm{d}V},
\end{equation}
where $f(\phi,c) = p(\phi) f^\alpha(c^\alpha) + (1-p(\phi)) f^\beta(c^\beta)$ is the Helmholtz free energy density, $f^\alpha$ and $f^\beta$ are the Helmholtz free energy densities of each phase, which are related to the grand-potential densities as $\omega^\alpha = f^\alpha - c^\alpha \tilde{\mu}^\alpha$. Equation~\ref{eq:F:binary} can be reformulated using Eq.~\ref{eq:GP:Omega:binary} as
\begin{equation}\label{eq:F:binary:2}
  F[\phi,c] = \Omega[\phi,\tilde{\mu}] + \intj{V}{c \tilde{\mu}}{V},
\end{equation}
where the local equilibrium constraint $\tilde{\mu}=\tilde{\mu}^\alpha=\tilde{\mu}^\beta$ is imposed. Mass conservation can also be introduced using a Lagrange multiplier, $\lambda$, resulting in the expression $\Omega + \lambda \int_V (c-c_0) \,\mathrm{d} V$, where $c_0$ represents the initial concentration. While this expression is similar to Eq.~\ref{eq:F:binary:2}, it is equivalent to the Helmholtz free energy only at equilibrium, where $\tilde{\mu}$ becomes constant and $\tilde{\mu}=\lambda$. These two methods lead to different dynamics, with our approach offering a self-consistent description of the dynamics.

The evolution of the Helmholtz free energy requires
\begin{align}
  \dod{F}{t}& = \dod{\Omega}{t} + \intj{V}{\left(\dpd{c}{t} \tilde{\mu} + c \dpd{\tilde{\mu}}{t}\right)}{V}\label{eq:dF}\\
            &= \intj{V}{\dfd{\Omega}{\phi}\dpd{\phi}{t} + \tilde{\mu} \dpd{c}{t}}{V}\label{eq:dF:2}\\
            &= \intj{V}{\dfd{\Omega}{\phi}\dpd{\phi}{t} + \vec{J}\cdot\grad{\tilde{\mu}}}{V} - \intj{\partial V}{\hat{n}\cdot(\tilde{\mu}\vec{J})}{A}\label{eq:dF:3} \\
            &= \intj{V}{- M_\phi \left(\dfd{\Omega}{\phi}\right)^2 - M \abs{\grad{\tilde{\mu}}}^2}{V} \leq 0,\label{eq:dF:4}
\end{align}
where Eq.~\ref{eq:GP:dOmegadt} is substituted into Eq.~\ref{eq:dF}, and Eq.~\ref{eq:GP:evolution:c:binary} is used in Eq.~\ref{eq:dF:2} with integration by parts. Equation~\ref{eq:dF:4} is obtained using the constitutive relations Eqs.~\ref{eq:GP:evolution:phi:binary} and \ref{eq:GP:evolution:J}, and the assumption that there is no mass flux across the boundary of $V$: $\hat{n}\cdot\vec{J}=0$, where $\hat{n}$ is the outward interfacial normal vector. Since $M_\phi$ and $M$ are both positive, Eq.~\ref{eq:dF:4} shows the evolution equations in the grand-potential phase field model [Eqs.~\ref{eq:GP:evolution:phi:binary} and \ref{eq:GP:evolution:c:binary}] minimizes the system's Helmholtz free energy [Eq.~\ref{eq:F:binary}].

In deriving Eq.~\ref{eq:dF:4}, we assume there is no mass flux at the boundary (closed system). For an open system that can exchange mass with the surroundings (or reservoir) at its boundary, a boundary integral is needed to account for this exchange process. The reservoir is assumed to be large compared to the system, so any exchange of mass does not alter its diffusion potential $\tilde{\mu}_{\text{resv}}$. The proper quantity to be minimized for this open system is
\begin{equation}\label{eq:F:L}
  L[\phi,c] = F[\phi,c] + \intj[t]{0}{\intj{\partial V}{\hat{n}\cdot (\tilde{\mu}_{\text{resv}} \vec{J}) }{A}}{t}.
\end{equation}
The time derivative of $L$ can be worked out similarly as
\begin{align}
\dod{L}{t} &= \intj{V}{\dfd{F}{\phi}\dpd{\phi}{t} + \vec{J}\cdot\grad{\tilde{\mu}}}{V} \nonumber\\
&- \intj{\partial V}{\hat{n}\cdot((\tilde{\mu}-\tilde{\mu}_{\text{resv}})\vec{J})}{A} \label{eq:dLdt1}\\
&= \intj{V}{- M_\phi \left(\dfd{F}{\phi}\right)^2 - M \abs{\grad{\tilde{\mu}}}^2}{V} \nonumber\\
&- \intj{\partial V}{M_t(\tilde{\mu}-\tilde{\mu}_{\text{resv}})^2}{A} \leq 0, \label{eq:dLdt2}
\end{align}
where we use constitutive relations Eqs.~\ref{eq:GP:evolution:phi:binary} and \ref{eq:GP:evolution:J} and boundary condition $\hat{n}\cdot\vec{J} = M_t (\tilde{\mu}-\tilde{\mu}_{\text{resv}})$, where $M_t$ is a trans-interface atomic mobility that is positive. In an electrochemical system, $\tilde{\mu}-\tilde{\mu}_{\text{resv}}$ can be related to the overpotential. If atoms jump across the boundary fast ($M_t=+\infty$), we have $\tilde{\mu}=\tilde{\mu}_{\text{resv}}$ at the boundary, indicating that the system exchanges mass with the surroundings at a constant $\tilde{\mu}$. This means $L$ is monotonically decreasing in time according to the evolution equations in Eqs.~\ref{eq:GP:evolution:phi:binary} and \ref{eq:GP:evolution:mu:binary}, while the grand potential $\Omega$ is not. At equilibrium, when mass diffusion has finished, the system has a uniform $\tilde{\mu}$ the same as the reservoir $\tilde{\mu}=\tilde{\mu}_{\text{resv}}$. $L$ can be written as $L= F - \tilde{\mu} N +\tilde{\mu}_{\text{resv}}N_0 = \Omega + \text{constant}$, where $N_0$ and $N$ are the total number of atoms in the system at $t=0$ and after equilibrium, respectively. So, both $L$ and $\Omega$ are at a minimum at equilibrium, which is consistent with classical thermodynamics. 

Now, we consider a timescale that is much larger than the diffusion timescale. This means that diffusion happens so quickly that it appears that the system is connected to a mass reservoir everywhere, maintaining a constant $\tilde{\mu}$. As a result, we have $\tilde{\mu}=\tilde{\mu}_{\text{resv}}$ and $\vec{J}=-M\grad{\tilde{\mu}}=\vec{0}$ at all points in the system. One example of such a system is active particles in batteries with infinite diffusion or small particle size that are in contact with an electrolyte reservoir \cite{Smith2017}. For this open system that can exchange mass with the surroundings everywhere, the proper quantity to be minimized is
\begin{equation}
  \label{eq:F:L2}
  L^*[\phi,c] = F[\phi,c] + \intj[t]{0}{\intj{V}{\tilde{\mu}_{\text{resv}}j }{V}}{t},
\end{equation}
where $j$ is a body mass flux from the system to the surroundings (sink). The continuity equation, Eq.~\ref{eq:GP:evolution:c:binary}, in this case, is $\partial c/\partial t = -j$, which can be substituted into Eq.~\ref{eq:F:L2} to give
\begin{align}
  L^*[\phi,c] &= F[\phi,c] - \intj{V}{c\tilde{\mu}}{V} + \intj{V}{c_0\tilde{\mu}_{\text{resv}}}{V}\nonumber\\
  &= \Omega[\phi,\tilde{\mu}] + \tilde{\mu}_{\text{resv}}N_0.\label{eq:F:L2:Omega}
\end{align}
where $c_0$ is the concentration at $t=0$. Note the last term in Eq.~\ref{eq:F:L2:Omega} is a constant. Therefore, $L^*$ is equivalent to the grand potential $\Omega$ at all times, while $L$ is equivalent to $\Omega$ only at equilibrium. It is critical to distinguish two cases: (i) the system is only connected to the reservoir at the system's outer boundary (minimizing $L$), and (ii) the system is connected to the reservoir at every point within the system or each subsystem is in contact with the reservoir (minimizing $\Omega$). The first case can be handled by a boundary integral, as in Eq.~\ref{eq:F:L}, while the second case requires the Legendre transformation of the free energy \textit{density}, as in Eq.~\ref{eq:F:L2:Omega}.

In summary, in a closed system (no mass moving across its boundary), the Helmholtz free energy in Eq.~\ref{eq:F:binary} is dissipated (decreases monotonically) as the system evolves. In an open system where mass can be exchanged with the surroundings only through its boundary, $L$ in Eq.~\ref{eq:F:L} is minimized. For an open system that can exchange mass with the surroundings throughout the system to maintain an iso-$\tilde{\mu}$ condition, the grand potential in Eq.~\ref{eq:GP:Omega:binary} or $L^*$ in Eq.~\ref{eq:F:L2} is minimized. In all cases, the core quantity to be minimized is the Helmholtz free energy. The interaction with the surroundings is taken into account through various ``boundary'' integrals, such as those in Eqs.~\ref{eq:F:L} and \ref{eq:F:L2}. Here, ``boundary'' refers to places where the system comes into contact with the reservoir. If we relax the isothermal assumption, the quantity to be minimized would be the negative entropy along with an appropriate boundary integral 
\cite{Sekerka2011, Penrose1990}. This aligns with the equivalence of thermodynamic ensembles in statistical mechanics, which states that the boundary conditions determine the proper ensemble to employ.\\

Plapp's grand-potential phase field model \cite{Plapp2011} was derived from the grand-potential functional in Eq.~\ref{eq:GP:Omega:binary}. Since we showed above that the Helmholtz free energy or $L$ is the proper quantity to be minimized, we here provide an alternative derivation of Plapp's model without utilizing the system's grand potential. Considering a system with $M$ phases and $N$ components, the independent variables are the phase field vector $\boldsymbol{\phi}=\{\phi_\alpha\}_{\alpha=1}^{M}$, and the concentration vector $\boldsymbol{c} = \{c_i\}_{i=1}^{N-1}$. Assuming all species have the same molar volume $V_m$, we have $\sum_{i=1}^N c_i = 1/V_m$, so there are $N-1$ independent concentrations. We choose the $N$th component to define the diffusion potential $\tilde{\mu}_i=\mu_i-\mu_N$, where $\mu_i$ is the chemical potential of species $i$. \\
The Helmholtz free energy of the system is
\begin{align}\label{eq:F}
F[\boldsymbol{\phi},\boldsymbol{c}]& = \int_{V}\! \sum_{\alpha=1}^{M}\frac{1}{2}\kappa \abs{\grad{\phi_\alpha}}^2 + m g(\boldsymbol{\phi}) \nonumber\\
    &\quad + \sum_{\alpha=1}^{M} p_\alpha(\boldsymbol{\phi}) f^\alpha(\boldsymbol{c}^\alpha(\tilde{\boldsymbol{\mu}}(\boldsymbol{\phi},\boldsymbol{c}))) {\,\mathrm{d}V}.
\end{align}
where $g(\boldsymbol{\phi})$ is a multiwell potential \cite{Fan1997}, $p_\alpha(\boldsymbol{\phi})$ is the interpolation function of phase $\alpha$ for the multiple-phase case \cite{Moelans2011}, $\boldsymbol{c}^\alpha=\{c_i^\alpha\}_{i=1}^{N-1}$, and $\tilde{\boldsymbol{\mu}}=\{\tilde{\mu}_i\}_{i=1}^{N-1}$. The key difference between our treatment and the KKS model (and its multiphase and multicomponent equivalent) is that we assume the phase concentrations $c_i^\alpha$ are a function of the diffusion potentials $\tilde{\mu}_i^\alpha$, which are in turn a function of the independent variables $\boldsymbol{\phi}$ and $\boldsymbol{c}$. This treatment allows us to directly impose the local equilibrium constraint $\tilde{\mu}_i = \tilde{\mu}_i^\alpha = \partial f^\alpha/\partial c_i^\alpha, \forall \alpha$. \\
The mixture rule for the phase concentrations is
\begin{equation}\label{eq:KKS:concentration:multiple}
  c_i = \sum_{\alpha=1}^{M} p_\alpha(\boldsymbol{\phi}) c_i^\alpha , \quad i=1,\cdots,N-1.
\end{equation}
The functional derivatives of $F$ to the independent field variables are derived using the chain rule and following the same procedure as in \cite{Moelans2011,Aagesen2018}
\begin{equation}\label{eq:KKS:dFdphi}
  \dfd{F}{\phi_\alpha} = - \kappa \lapl{\phi_\alpha} + m \dpd{g(\boldsymbol{\phi})}{\phi_\alpha} + \sum_{\beta=1}^M \dpd{p_\beta(\boldsymbol{\phi})}{\phi_\alpha} \omega^\beta,
\end{equation}
\begin{equation}\label{eq:KKS:dFdc}
  \dfd{F}{c_i} = \tilde{\mu}_i.
\end{equation}
The grand potential density $\omega$ in Eq.~\ref{eq:KKS:dFdphi} arises naturally from the local equilibrium constraint and Eq.~\ref{eq:KKS:concentration:multiple}.\\
The evolution equations that minimize the system's Helmholtz free energy [Eq.~\ref{eq:F}] are (neglecting cross-coupling between $\boldsymbol{\phi}$ and $\boldsymbol{c}$)
\begin{equation}
  \label{eq:mGP:dphidt}
  \dpd{\phi_\alpha}{t}  = - M_\phi^\alpha \dfd{F}{\phi_\alpha} ,
\end{equation}
\begin{equation}
  \label{eq:mGP:dcdt}
  \dpd{c_i}{t}  =  \divj{\left(\sum_{j=1}^{N-1} M_{ij} \grad{\tilde{\mu}_j}\right)},
\end{equation}
where $M_\phi^\alpha$ are the phase field mobilities and $M_{ij}$ are the atomic mobilities. At this point, we still need to solve the local equilibrium constraint $\tilde{\mu}_i=\tilde{\mu}_i^1=\cdots=\tilde{\mu}_i^M, i=1,\cdots,N-1$ at each grid point and time step to get the phase concentrations $c_i^\alpha$. This problem can be resolved by reformulating the mass conservation equations [Eq.~\ref{eq:mGP:dcdt}] using the diffusion potentials $\tilde{\mu}_i$, a critical step in the original grand-potential phase field model \cite{Plapp2011}, as
\begin{align}
  \label{eq:mGP:dmudt}
  \chi_{ij}\dpd{\tilde{\mu}_j}{t} =& - \divj{\left(\sum_{j=1}^{N-1} M_{ij} \grad{\tilde{\mu}_j}\right)} \nonumber\\
  &- \sum_{\alpha=1}^{M}\left(\sum_{\beta=1}^{M} \dpd{p_\beta(\boldsymbol{\phi})}{\phi_\alpha} c_i^\beta\right) \dpd{\phi_\alpha}{t},
\end{align}
where the susceptibility is $\chi_{ij}=\partial c_i/\partial \tilde{\mu}_j$.\\
The phase concentrations $c_i^\alpha$ can then be solved from the diffusion potentials $\tilde{\mu}_i$, and there is no need to explicitly impose the local equilibrium constraint like the KKS model. Equation~\ref{eq:mGP:dmudt} should not be viewed as the evolution of the independent field variables but as a reformulation of the mass conservation equations in Eq.~\ref{eq:mGP:dcdt}. The independent field variables remain the concentrations since the Helmholtz free energy is minimized. Notably, when Eq.~\ref{eq:mGP:dmudt} is solved numerically, it can lead to an accumulated error in mass conservation, which requires careful treatment \cite{Khanna2020}.

This work addresses an apparently widespread misapprehension regarding the grand-potential phase field model: What thermodynamic potential is minimized or dissipated during evolution? Although the original model's equations are correct, they do not result in a monotonically decreasing grand potential. Instead, the thermodynamic potential to be minimized is the Helmholtz free energy with the proper boundary condition. We identified a key difference between minimizing the Helmholtz free energy and the grand potential, which is related to how the system exchanges mass with the surroundings. When the system can only exchange mass with the surroundings through its outer boundary, the Helmholtz free energy (with boundary condition as boundary integral) decreases monotonically with time. On the other hand, if the system can exchange mass locally everywhere within the system, a finite diffusion rate is not present, and the grand potential is the proper quantity to be minimized. Although both quantities are at a minimum at equilibrium, they lead to different evolution equations, corresponding to physical processes at different timescales.

We acknowledge the financial assistance Award 70NANB14H012 from the U.S. Department of Commerce, National Institute of Standards and Technology as part of the Center for Hierarchical Materials Design (CHiMaD). We would like to thank Professor Mathis Plapp for his valuable comments.

%

\end{document}